\documentclass[twocolumn,secnumarabic,amssymb,nobibnotes, aps, prapplied,10pt]{revtex4-2}

\usepackage{mathrsfs,amsmath,bm,graphicx}
\usepackage{dcolumn}  
\renewcommand{\arraystretch}{2}
\usepackage{verbatim}
\usepackage{upgreek}
\usepackage{xcolor}
\begin{document}


\title{Far-field and near-field directionality in acoustic scattering}


\author{Lei Wei}
\email{lei.wei@kcl.ac.uk}
\affiliation{Department of Physics, King's College London, Strand, London, WC2R 2LS, United Kingdom}
\author{Francisco J. Rodr\'{i}guez-Fortu\~{n}o}
\email{francisco.rodriguez_fortuno@kcl.ac.uk}
\affiliation{Department of Physics, King's College London, Strand, London, WC2R 2LS, United Kingdom}


\date{\today}

\begin{abstract}
Far-field directional scattering and near-field directional coupling from simple sources have recently received great attention in photonics: beyond circularly-polarized dipoles, whose directional coupling to evanescent waves was recently applied to acoustics, the near-field directionality of modes in optics includes phased combinations of electric and magnetic dipoles, such as the Janus dipole and the Huygens dipole, both of which have been experimentally implemented using high refractive index nanoparticles. In this work we extend this to acoustics: we propose the use of high acoustic index scatterers exhibiting phased combinations of acoustic monopoles and dipoles with far-field and near-field directionality. All solutions stem from the elegant acoustic angular spectrum of the acoustic source, in close analogy to electromagnetism. A Huygens acoustic source with zero backward scattering is proposed and numerically demonstrated, as well as a Janus source achieving face-selective and position-dependent evanescent coupling to nearby acoustic waveguides.
\end{abstract}

\maketitle
\section{Introduction}

In electromagnetism and photonics, high index dielectric particles are becoming an important platform to study novel physical phenomena \cite{AIKuznetsovSci2016}. Unlike plasmonic nanoparticles, a high index dielectric particle can exhibit strong magnetic Mie resonances \cite{AGarciaEtxarriOE2011,AIKuznetsov2012} that are of comparable strength to the electric ones. Sources such as the Huygens and Janus dipoles show interesting directional scattering and coupling characteristics, both in the far field and in the near field \cite{MFPicardiPRL2018,LWeiPRAppl2020}, and they have been experimentally demonstrated in high-index nanoparticles \cite{JMGeffrinNC2012,YHFuNC2013,SPersonNL2013,MFPicardiLSA2019} with interfering electric $\mathbf{p}$ and magnetic $\mathbf{m}$ dipole moments. In the far-field, the combination of orthogonal $\mathbf{p}$ and $\mathbf{m}$ dipoles following Kerker's condition $p=m/c$ gives rise to the Huygens dipole, exhibiting directional scattering \cite{JMGeffrinNC2012,YHFuNC2013,SPersonNL2013} with applications in reflectionless metasurfaces \cite{IStaudeACSNano2013,CPfeifferPRL2013,AIKuznetsovSci2016} and optical metrology \cite{LWeiPRL2018} among others. In the near-field, directional coupling of waveguided modes was initially predicted and demonstrated in electromagnetism via the evanescent coupling of circularly polarised dipoles \cite{FJRFortunoSci2013}, relying on the transverse spin and spin-momentum locking in evanescent waves \cite{FJRFortunoNPhoton2015,ZJacobOptica2016,KYBliokhSci2015}. The analog acoustic scenario was recently demonstrated using circular acoustic dipoles \cite{YLongPNAS2018, CShiNSR2019}. This shows that the transverse spin is a universal property of evanescent waves in any wave field including acoustics \cite{KYBliokhPRB2018R, IDToftulPRL2019,YLongPNAS2018, CShiNSR2019,LBurnsArXiv2019}, electromagnetism \cite{FJRFortunoNPhoton2015,ZJacobOptica2016,KYBliokhSci2015} and gravitational waves \cite{SGolatArxiv2019}. However, near-field directionality in electromagnetism was generalised beyond circular dipoles to include combinations of electric and magnetic dipoles that achieve near-field directional coupling with different symmetries \cite{MFPicardiPRL2018,LWeiPRAppl2020}: one example is the aforementioned Huygens dipole, which can be also applied for near-field directionality, and another is the intriguing Janus dipole, whose combination between electric and magnetic dipoles requires a 90 degree phase difference to achieve a face-dependent or position-dependent coupling to the waveguide modes \cite{MFPicardiPRL2018}. These sources exploit the amplitude and phase relations that exist between different components of the electric and magnetic fields in evanescent waves. In acoustics, similar amplitude and phase relations exist between the scalar pressure and vector velocity fields, opening the possibility of Huygens and Janus-like directional sources.

High index materials are also sought after in acoustics. Micro-sized air bubbles in liquid show strong resonances \cite{MKafesakiPRL2000} and are widely used as a contrast agent for high resolution acoustic imaging \cite{CErricoNature2005}. Acoustic metamaterials \cite{SACummerNatureRM2016,FZangenehNejadRevPhys2019} made of high index materials including air bubbles and porous silicone rubbers \cite{ABaSR2017} are proposed to achieve exotic physical properties like negative effective mass density and modulus \cite{JensenLiPRE2004,YDingPRL2007}. Mie-type acoustic meta-atoms \cite{JJordaanAPL2018,GLuAPL2017,YChengNatMat2015} have been proposed and demonstrated, which can have high effective acoustic index even in the background of air. In this work we explore the possibility to produce acoustic Huygens and Janus-type directional sources to achieve far and near-field directionality, using a high-index particle platform.

\section{Theory}

We begin this work by deriving all possible combinations of an acoustic monopole $M$ and dipole $\mathbf{D}$ that achieve far- and near-field directionality. The complex pressure field of such a source is given by:

\begin{align}\label{eq:MDpressure}
p(\mathbf{r})&=M \frac{\mathrm{e}^{ik_0r}}{k_0r} + \frac{1}{ik_0}\mathbf{D}\cdot\nabla \left(\frac{\mathrm{e}^{ik_0r}}{k_0r}\right),
\end{align}

\noindent where $r=|\mathbf{r}|$ is the distance to the source, assumed to be at the origin, and $k_0 = 2\pi/\lambda$ is the acoustic wave-number of free space. To analyse both far- and near-field directionality we will apply a standard technique in electromagnetism: the angular spectrum decomposition \cite{LNovotnyJOSAA1997, MFPicardiPRB2017,LMandelBOOK,MNietoVesperinasBOOK}. Such decomposition expands the fields as a superposition of momentum eigenmodes $p(\mathbf{r})=\int_\mathbf{k} p(\mathbf{k}) e^{i \mathbf{k} \cdot \mathbf{r}} \mathrm{d}\mathbf{k}$. Each component $p(\mathbf{k}) e^{i \mathbf{k} \cdot \mathbf{r}}$ has a constant wave-vector $\mathbf{k} = (k_x,k_y,k_z)$. Owing to the dispersion relation, the $k_z$ component of $\mathbf{k}$ can be derived from the in-plane momentum $(k_x,k_y)$ via the dispersion relation $k_x^2+k_y^2+k_z^2=k_0^2$. As is well-known in photonics, in the region $k_x^2+k_y^2\leq k_0^2$ the momentum eigenmodes correspond to propagating plane waves with a real-valued $\mathbf{k}$. However, in the region $k_x^2+k_y^2 > k_0^2$, the component $k_z$ becomes imaginary, and $e^{i \mathbf{k} \cdot \mathbf{r}}$ represents an evanescent wave, corresponding to the near-field spectrum \cite{JDMaynardJASA1985}.

\begin{center}
\begin{figure}[!htp]
\includegraphics[width=0.7\textwidth]{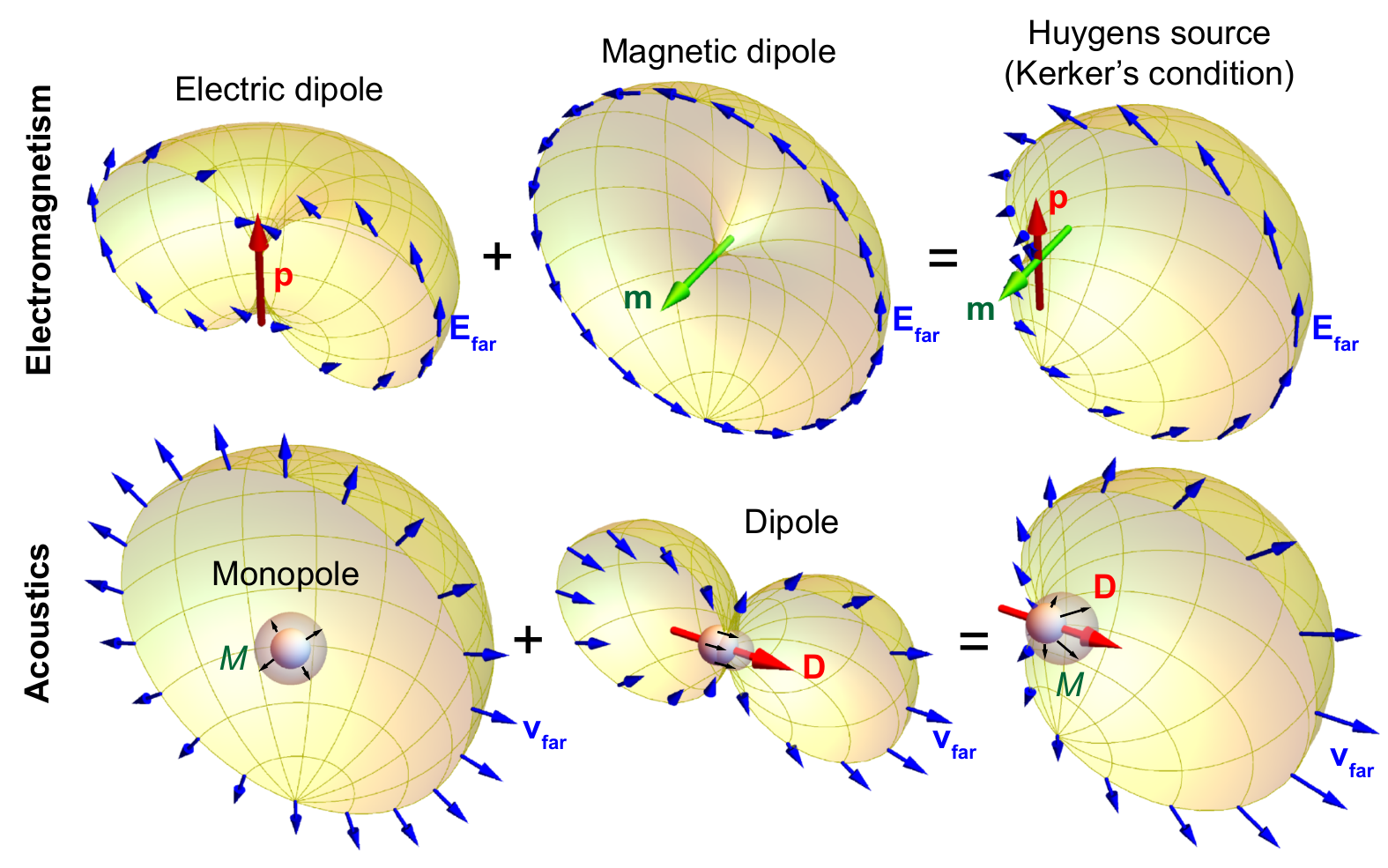}
\caption{Analogy between far field directionality (Kerker's condition) in electromagnetic and acoustic scattering.}
\label{fig1l}
\end{figure}
\end{center}

The angular spectrum $p(\mathbf{k})$ can be analytically calculated via a partial Fourier transform of $p(\mathbf{r})$ from Eq.\ \ref{eq:MDpressure}, using Weyl's identity \cite{LMandelBOOK}, and it is given as (see supplementary information):

\begin{align}\label{eq:MDpressurefourier}
p(\mathbf{k})&= \frac{i}{2\pi k_0 k_z}\left( M + \hat{\mathbf{k}}\cdot\mathbf{D} \right),
\end{align}

\noindent where $\hat{\mathbf{k}} = \mathbf{k}/k_0$. Eq.\ \ref{eq:MDpressurefourier} is the master equation from which any type of directionality can be analysed or designed. Far-field directionality manifests itself as zeroes in the angular spectrum inside the circle $k_x^2+k_y^2=k_0^2$, while near-field directionality manifests itself as zeroes outside of that circle \cite{MFPicardiPRL2018,LWeiPRAppl2020,MFPicardiLPR2019}. For example, let's start with far-field directionality: to achieve directionality in the forward $x$ direction we may introduce a zero of $p(\mathbf{k})$ for the plane wave propagating along the negative $x$ axis. Substituting $\hat{\mathbf{k}}=(-1,0,0)$ into Eq.\ \ref{eq:MDpressurefourier} and equating it to zero, one immediately arrives at the acoustic analogue of Kerker's condition $M - D_x = 0$. An acoustic monopole $M$ combined with an acoustic dipole $\mathbf{D}=(M,0,0)$ will result in Kerker-like far-field directionality, in complete analogy to a Huygens' dipole. This is shown in the far-field diagrams of Fig.\ \ref{fig1l}. Intuitively, the monopole source is expanding and contracting in an oscillating manner, creating an isotropic spherical pressure wave, while the dipolar source is vibrating back and forth, creating a peanut-like radiation diagram, with opposite pressure changes and opposite velocities on opposite directions. Their coherent combination results in a very special vibration of the source: the source moves forwards while expanding, and then moves backwards while contracting, in such a way that the backward-facing surface does not move, producing no pressure wave in the backward direction. In the next section we show how to implement this acoustic Huygens source in a realistic spherical or cylindrical scatterer upon plane wave excitation, exhibiting no back-scattering, with interesting applications.

\onecolumngrid
\widetext
\twocolumngrid

\begin{figure}[!htp]
\centering
\includegraphics[width=0.44\textwidth]{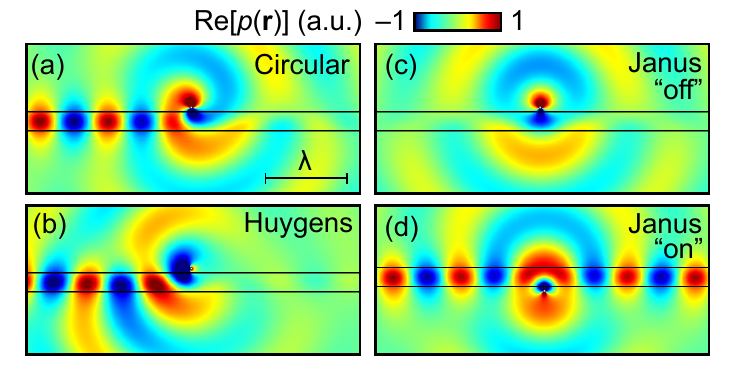}
\caption{Near-field directional coupling using acoustic monopole and dipole combinations. (a) Circular dipole, (b) Huygens source, (c-d) Janus source. The waveguide slab has $\bar{\rho}=\bar{\beta}=2$, thickness $0.2 \lambda$, and $n_\mathrm{eff}\approx1.31$. The source is placed at a distance $0.05 \lambda$ above (a-c) or below (d) the waveguide.}
\label{fig_sourceswg}
\end{figure}

Even more interesting solutions appear if we look at near-field directionality. In this case, we must set the angular spectrum in Eq.\ \ref{eq:MDpressurefourier} to be zero at some value of $(k_x,k_y)$ outside of the circle $k_x^2+k_y^2=k_0^2$. Following an identical approach to the optical case \cite{MFPicardiPRB2017,MFPicardiLPR2019, LWeiPRAppl2020}, we can study near-field directional coupling of a waveguided mode with an effective refractive index $n_\mathrm{eff}$. The evanescent wave near-field component that would couple to such a mode, propagating in the $\pm x$ direction, is given by $\hat{\mathbf{k}}=(\pm n_\mathrm{eff},0,i \gamma)$, where $\gamma = \pm \sqrt{n_\mathrm{eff}^2 - 1}$ to fulfill the wave-equation condition $\hat{\mathbf{k}}\cdot\hat{\mathbf{k}}=1$. The sign of $\pm n_\mathrm{eff}$ will determine the direction of propagation of the mode, $+x$ or $-x$, while the sign of $\gamma$ will determine the position of the source, below or above the waveguide, respectively. Substituting this $\hat{\mathbf{k}}$ into Eq.\ \ref{eq:MDpressurefourier}, and equating it to zero, we immediately arrive at $M+n_\mathrm{eff} D_x + i \gamma D_z = 0$. Three simple solutions emerge when only two of the three source components are allowed to be non-zero: (i) the circularly polarized dipole, (ii) the near-field Huygens dipole, and (iii) the Janus dipole. The three solutions are summarized in Table \ref{tab:table1} and simulated numerically in Comsol by placing the different sources near a waveguide, shown in Fig.\ \ref{fig_sourceswg}. The sources are a clear mathematical analogy to their electromagnetic counterparts \cite{MFPicardiPRL2018}. While in electromagnetism we could find two versions of each solution --corresponding to each of the two transverse polarizations--, in acoustics there is only one version of each solution, consistent with the fact that acoustic waves have a single longitudinal polarisation. In the next section we show how high acoustic index particles can be used to achieve these solutions, with the required relative amplitudes and phases between the monopole and dipole components, and numerically demonstrate Huygens and Janus behaviour in the far-field and near-field, respectively.

\begingroup
\begin{table}
\begin{center}
\caption{Elemental monopole and dipole combinations for near-field directionality in planar waveguides.} 
\label{tab:table1}
\begin{tabular}{ |c|c|c| } 
\hline
  Source & Condition \\
\hline
Circular & $D_x = \frac{-i \sqrt{n_\mathrm{eff}^2 - 1}}{n_\mathrm{eff}} D_z$ \\ 
Huygens & $M=-n_\mathrm{eff} D_x$ \\ 
Janus & $M=-i \sqrt{n_\mathrm{eff}^2 - 1} D_z$ \\ 
\hline
\end{tabular}
\end{center}
\end{table}
\endgroup

\section{High index acoustic scatterers}

Consider a high index acoustic scatterer upon which an external, time-harmonic sound wave with pressure distribution $p_{\mathrm{in}}(\mathbf{r})$ and velocity field $\mathbf{v}_{\mathrm{in}}=\frac{1}{i\omega\rho_0}\nabla p_{\mathrm{in}}$ is incident. We assume the scatterer is located at $\mathbf{r}=0$ in a background with mass density $\rho_0$ and compressibility $\beta_0$ and only longitudinal sound waves with velocity $c_0=1/\sqrt{\rho_0\beta_0}$ considered. The monopole and dipole induced in the acoustic scatterer are given by:

\begin{align}
M=\alpha_{M} p_{\mathrm{in}}, \quad \quad
\mathbf{D}=\alpha_{D} \sqrt{\frac{\rho_0}{\beta_0}}\mathbf{v}_{\mathrm{in}},
\end{align}\label{eq:MDmoments}


\noindent where $p_{\mathrm{in}}$ and $\mathbf{v}_{\mathrm{in}}$ are evaluated at $\mathbf{r}=0$, and $\alpha_{M}$ and $\alpha_{D}$ represent the acoustic monopolar and dipolar strength, solely determined by the scatterer and the background material. In the special case of plane wave or evanescent wave incidence $p_{\mathrm{in}}(\mathbf{r})= p_0 \mathrm{e}^{ik_0\mathbf{\hat{k}_{in}\cdot\mathbf{r}}}$, the dipole moment is reduced to $\mathbf{D}=\mathbf{\hat{k}_{in}}\alpha_{D}p_0$ and the master equation for the angular spectrum of the scattered field, Eq.\ \ref{eq:MDpressurefourier}, can be simplified as:

\begin{align}\label{eq:MDpressurefourierplanewave}
p(\mathbf{k})&= \frac{ip_0}{2\pi k_0 k_z}\left[ \alpha_{\mathrm{M}} +  \alpha_{\mathrm{D}}\left(\hat{\mathbf{k}}\cdot\mathbf{\hat{k}}_\mathrm{in}\right) \right].
\end{align}

In order to illustrate our concept in a simple manner but without loss of generality, let's assume the scatterer is a sphere of radius $r_0$ and made of a material that supports longitudinal sound waves only, and has a relative mass density and compressibility $\bar{\rho}=\rho_1/\rho_0$ and $\bar{\beta}=\beta_1/\beta_0$. The acoustic scattering of spheres and cylinders can be analytically calculated (as detailed in the supplementary) in a similar way to Mie theory for optical scattering. A high acoustic index $n=\sqrt{\bar{\rho}\bar{\beta}}$ corresponds to a strong contrast in the speed of sound between the scatterer and the background medium $c_1=c_0/n$. Just like the electromagnetic case, where a high refractive index results in a spectral region (i.e. certain values of $2\pi r_0/\lambda$) dominated by the electric and magnetic dipolar contribution, a high index in acoustics also results in a spectral region of strong acoustic monopolar and dipolar responses, with the higher order modes suppressed.

\begin{figure}[!htp]
\centering
\includegraphics[]{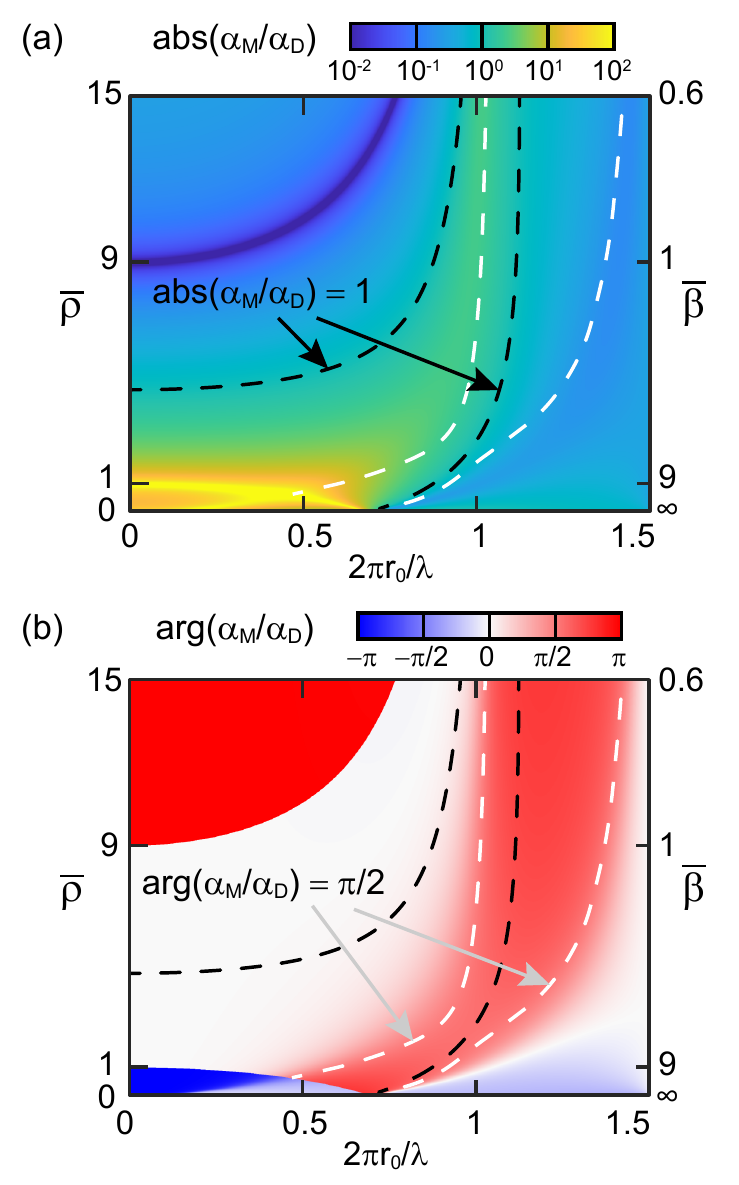}
\caption{Relative amplitude (a) and phase (b) of the acoustic monopolar and dipolar moments $\alpha_{\mathrm{M}}$ and $\alpha_{\mathrm{D}}$ for a sphere with acoustic index $n=3$. The black dashed lines indicate the parametric locations where $|\alpha_{\mathrm{M}}/\alpha_{\mathrm{D}}|=1$, and the white dashed lines indicate the parametric locations where $\mathrm{arg}\{\alpha_{\mathrm{M}}/\alpha_{\mathrm{D}}\}=\pi/2$.}
\label{fig2l}
\end{figure}

The relative amplitude and phase between the monopolar and dipolar strength can be tuned with the material properties and size, enabling us to easily achieve the specific conditions required for Huygens and Janus sources. Fig.\ \ref{fig2l} shows the relative amplitude and phase of the acoustic monopolar and dipolar moments for a sphere with an acoustic index $n=3$, with varying relative mass density $\bar{\rho}$ and compressibility $\bar{\beta}$. In the range shown, $0<2\pi r_0/\lambda<1.5$, the higher order multipoles are negligible. We begin by looking at the conditions required for a Huygens-type far-field directional particle. Following Eq.\ \ref{eq:MDpressurefourierplanewave}, the scattering pressure of the particle in the forward/backward direction, relative to the incident plane wave, is given by $p(\mathbf{\pm\mathbf{\hat{k}}_\mathrm{in}}k_0)\propto \alpha_{\mathrm{M}} +  \alpha_{\mathrm{D}}\left[(\pm\mathbf{\hat{k}}_\mathrm{in})\cdot\mathbf{\hat{k}}_\mathrm{in}\right]$. Owing to the dispersion relation, we know that $\mathbf{\hat{k}}_\mathrm{in}\cdot\mathbf{\hat{k}}_\mathrm{in}=1$, and so, the condition to achieve zero forward/backward scattering becomes:

\begin{align}\label{eq:MDpressurefourierhuygens}
p(\mathbf{\pm\mathbf{\hat{k}}_\mathrm{in}}k_0)&= \frac{ip_0}{2\pi k_0 k_z}\left[ \alpha_{\mathrm{M}} \pm  \alpha_{\mathrm{D}} \right]=0.
\end{align}

\begin{figure}[!htp]
\centering
\includegraphics[]{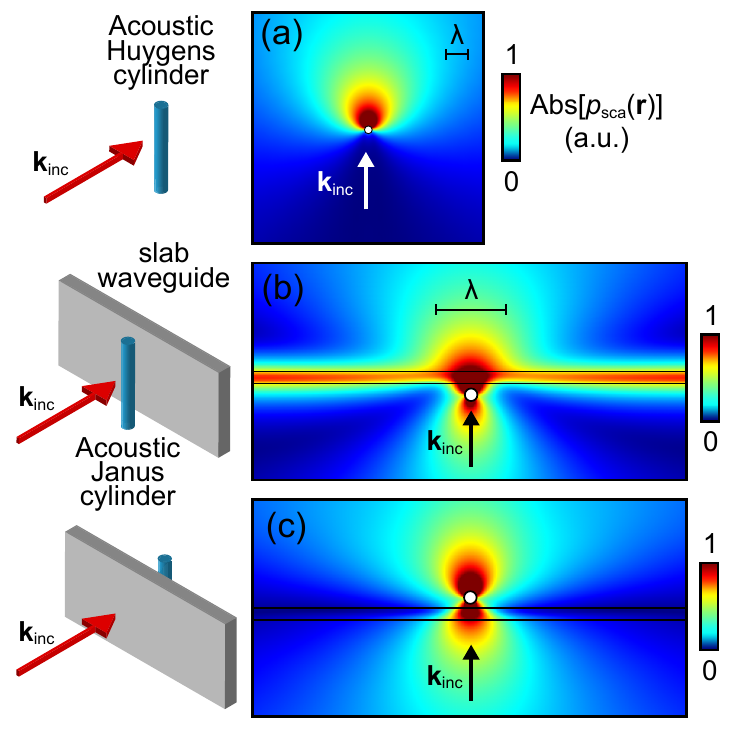}
\caption{(a) Scattered pressure distribution of a cylinder (radius $r_0=0.2\lambda/(2\pi)$, relative acoustic index $n=3$ and relative mass density $\bar{\rho}=4.2$) upon plane wave incidence along the $z$ direction, where the Huygens condition $\alpha_{\mathrm{M}}=\alpha_{\mathrm{D}}$ for zero backward scattering is met. (b, c) Scattered pressure distribution of a cylinder (radius $r_0=0.287\lambda/(2\pi)$, relative acoustic index $n=8.082$ and relative mass density $\bar{\rho}=1.04$) placed at a distance of $\lambda/8$ below (in (b)) or above (in (c)) a slab waveguide (thickness $\lambda/6$, relative acoustic index $n=1.81$ and relative mass density $\bar{\rho}=2$). A sound plane wave is incident along the $z$ direction. The cylinder scatterer fulfils the Janus condition with $\alpha_{\mathrm{M}}/\alpha_{\mathrm{D}}=0.599i$ and the slab waveguide supports a guided mode with effective index $n_{\mathrm{eff}}\approx1.166$.
}
\label{fig3l}
\end{figure}

The condition for no backward scattering is therefore $\alpha_{\mathrm{M}} = \alpha_{\mathrm{D}}$, and it can be easily implemented with a subwavelength particle. Consider the behavior shown in Fig.\ \ref{fig2l}(b) in the limit of small particles ($r_0/\lambda \to 0$). In this limit, we can see three distinct regions: the monopolar and dipolar moments are $\pm\pi$ out of phase (preferred backward scattering) in the ranges $0<\bar{\beta}<1$ and $0<\bar{\rho}<1$, while they are in phase (preferred forward scattering) in the overlapping region where both $\bar{\beta}$ and $\bar{\rho}$ are larger than 1. This result can be derived analytically: in the limit $r_0/\lambda \to0$, the two moments can be approximated as $\alpha_{\mathrm{M}}\approx(\bar{\beta}-1)(k_0r_0)^3/3$ and $\alpha_{\mathrm{D}}\approx(\bar{\rho}-1)(k_0r_0)^3/(2\bar{\rho}+1)$. In the region where both monopole and dipole are in phase, there is a point where they also have equal amplitudes, marked with a black dashed line, and so $\alpha_M=\alpha_D$. This condition represents the acoustic analog of the electromagnetic Huygens dipole with zero back-scattering in the far field, easily achievable with high-index spheres or cylinders. An acoustic Comsol simulation of such a particle (in the simpler two-dimensional case of a cylindrical scatterer) is shown in Fig.\ \ref{fig3l}(a), clearly showing the forward scattering.

The Janus condition can also be achieved, but not in the limit of small particles. The white dashed lines in Fig.\ \ref{fig2l} indicate the locations where the monopole and dipole moments are on quadrature phase difference $\alpha_{\mathrm{M}}/\alpha_{\mathrm{D}}=i|\alpha_{\mathrm{M}}/\alpha_{\mathrm{D}}|$ as required by Janus sources. The specific required amplitude ratio $|\alpha_{\mathrm{M}}/\alpha_{\mathrm{D}}|$ will depend on the effective index of the modes we want to couple to. Assume that the incident sound wave is propagating in the $+\hat{\mathbf{z}}$ direction, i.e. $\mathbf{\hat{k}}_\mathrm{in}=+\hat{\mathbf{z}}$. Following Eq.\ (\ref{eq:MDpressurefourierplanewave}), the scattering angular spectrum for evanescent wavevectors $\mathbf{k}_\mathrm{ev}$ with $(k_x^2+k_y^2)^{1/2}=k_0 n_\mathrm{eff} >k_0$, and corresponding $k_z=\pm ik_0\gamma$, is given by:

\begin{align}\label{eq:MDpressurefourierjanus}
p(\mathbf{k}_\mathrm{ev})&= \frac{ip_0}{2\pi k_0 k_z}\left[i \left|\frac{\alpha_{\mathrm{M}}}{\alpha_{\mathrm{D}}}\right| \pm  i\gamma \right],
\end{align}

\noindent where we introduced the Janus condition $\alpha_{\mathrm{M}}/\alpha_{\mathrm{D}}=i|\alpha_{\mathrm{M}}/\alpha_{\mathrm{D}}|$, and where $\gamma=(n_\mathrm{eff}^2-1)^{1/2}$ is always taken as the positive square root, while the $\pm$ sign corresponds to scattering in the $z>0$ or $z<0$ half spaces, respectively. It follows from Eq.\ \ref{eq:MDpressurefourierjanus} that the evanescent components with a transverse wavevector satisfying $(k_x^2+k_y^2)^{1/2}=k_0 n_\mathrm{eff}=k_0(1+|\alpha_{\mathrm{M}}/\alpha_{\mathrm{D}}|^2)^{1/2}$, will have a zero amplitude in the lower half space ($z<0$), meaning that the source will not couple to waveguided modes with an effective index $n_\mathrm{eff}$ when the waveguide is placed in the $z<0$ half-space. At the same time, the source has a non-zero amplitude for the same evanescent waves in the upper half space ($z>0$), meaning that it will couple to the modes of the same waveguide if it is placed in the $z>0$ half-space. This, together with no far-field directionality --which is easy to prove for these monopole and dipole compositions $\alpha_{\mathrm{M}}/\alpha_{\mathrm{D}}=i|\alpha_{\mathrm{M}}/\alpha_{\mathrm{D}}|$-- constitutes the signature of a Janus source, with its charcteristic face-dependent behaviour, in perfect analogy to electromagnetism. Fig.\ \ref{fig3l}(b-c) shows a cylindrical particle designed in this way, simulated in Comsol, and placed at either side of a planar slab, showing the side-dependent on-off coupling to the waveguide modes.

\section{Conclusion}
We have extended the analogy of far and near-field directionality from dipolar sources in electromagnetism to the domain of acoustics. On one hand, we theoretically describe and numerically demonstrate the acoustic analogy of a far-field directional Huygens dipole with zero back-scattering, implemented with high acoustic index spheres or cylinders. This may have interesting application for reduced reflection materials in acoustics engineering. The possibility of zero forward-scattering Huygens scatterers could also lead to the design of novel sound barriers.

On the other hand, we also theoretically described near-field directional coupling of waveguided modes using monopole and dipole combinations. We show that the three solutions: circular dipole, Huygens source, and Janus source, in perfect analogy to the electromagnetic case, appear naturally as independent solutions to the simple angular spectrum of an acoustic source. We theoretically and numerically propose a simple realistic way of achieving a Janus scatterer with spherical or cylindrical high acoustic index materials, clearly exhibiting the characteristic position-dependent near-field coupling behaviour. This has clear implications for the understanding and control of the near-fields of sound waves.

During the writing of this manuscript we noticed the recent publication of the theoretical and experimental Ref.\ \cite{YLongNSR2020} which describes and experimentally demonstrates the same set of three acoustic near-field directional sources described here. The work elegantly derives the three sources from fundamental symmetry considerations and follows a Fermi Golden rule approach. The key difference of our work is our angular spectrum approach and our proposed realization of the sources using the scattering from spherical or cylindrical particles made of high acoustic index materials, instead of phased combinations of acoustic monopoles.

\section*{Acknowledgement}
This work is supported by European Research Council Starting Grant No. ERC-2016-STG-714151-PSINFONI.
\bibliography{shout}

\begin{thebibliography}{39}%
\makeatletter
\providecommand \@ifxundefined [1]{%
 \@ifx{#1\undefined}
}%
\providecommand \@ifnum [1]{%
 \ifnum #1\expandafter \@firstoftwo
 \else \expandafter \@secondoftwo
 \fi
}%
\providecommand \@ifx [1]{%
 \ifx #1\expandafter \@firstoftwo
 \else \expandafter \@secondoftwo
 \fi
}%
\providecommand \natexlab [1]{#1}%
\providecommand \enquote  [1]{``#1''}%
\providecommand \bibnamefont  [1]{#1}%
\providecommand \bibfnamefont [1]{#1}%
\providecommand \citenamefont [1]{#1}%
\providecommand \href@noop [0]{\@secondoftwo}%
\providecommand \href [0]{\begingroup \@sanitize@url \@href}%
\providecommand \@href[1]{\@@startlink{#1}\@@href}%
\providecommand \@@href[1]{\endgroup#1\@@endlink}%
\providecommand \@sanitize@url [0]{\catcode `\\12\catcode `\$12\catcode
  `\&12\catcode `\#12\catcode `\^12\catcode `\_12\catcode `\%12\relax}%
\providecommand \@@startlink[1]{}%
\providecommand \@@endlink[0]{}%
\providecommand \url  [0]{\begingroup\@sanitize@url \@url }%
\providecommand \@url [1]{\endgroup\@href {#1}{\urlprefix }}%
\providecommand \urlprefix  [0]{URL }%
\providecommand \Eprint [0]{\href }%
\providecommand \doibase [0]{https://doi.org/}%
\providecommand \selectlanguage [0]{\@gobble}%
\providecommand \bibinfo  [0]{\@secondoftwo}%
\providecommand \bibfield  [0]{\@secondoftwo}%
\providecommand \translation [1]{[#1]}%
\providecommand \BibitemOpen [0]{}%
\providecommand \bibitemStop [0]{}%
\providecommand \bibitemNoStop [0]{.\EOS\space}%
\providecommand \EOS [0]{\spacefactor3000\relax}%
\providecommand \BibitemShut  [1]{\csname bibitem#1\endcsname}%
\let\auto@bib@innerbib\@empty
\bibitem [{\citenamefont {Kuznetsov}\ \emph {et~al.}(2016)\citenamefont
  {Kuznetsov}, \citenamefont {Miroshnichenko}, \citenamefont {Brongersma},
  \citenamefont {Kivshar},\ and\ \citenamefont
  {Luk'yanchuk}}]{AIKuznetsovSci2016}%
  \BibitemOpen
  \bibfield  {author} {\bibinfo {author} {\bibfnamefont {A.~I.}\ \bibnamefont
  {Kuznetsov}}, \bibinfo {author} {\bibfnamefont {A.~E.}\ \bibnamefont
  {Miroshnichenko}}, \bibinfo {author} {\bibfnamefont {M.~L.}\ \bibnamefont
  {Brongersma}}, \bibinfo {author} {\bibfnamefont {Y.~S.}\ \bibnamefont
  {Kivshar}},\ and\ \bibinfo {author} {\bibfnamefont {B.}~\bibnamefont
  {Luk'yanchuk}},\ }\bibfield  {title} {\bibinfo {title} {{Optically resonant
  dielectric nanostructures}},\ }\href@noop {} {\bibfield  {journal} {\bibinfo
  {journal} {Science}\ }\textbf {\bibinfo {volume} {354}},\ \bibinfo {pages}
  {aag2472} (\bibinfo {year} {2016})}\BibitemShut {NoStop}%
\bibitem [{\citenamefont {Garc\'{i}a-Etxarri}\ \emph
  {et~al.}(2011)\citenamefont {Garc\'{i}a-Etxarri}, \citenamefont
  {G\'{o}mez-Medina}, \citenamefont {Froufe-P\'{e}rez}, \citenamefont
  {L\'{o}pez}, \citenamefont {Chantada}, \citenamefont {Scheffold},
  \citenamefont {Aizpurua}, \citenamefont {Nieto-Vesperinas},\ and\
  \citenamefont {S\'{a}enz}}]{AGarciaEtxarriOE2011}%
  \BibitemOpen
  \bibfield  {author} {\bibinfo {author} {\bibfnamefont {A.}~\bibnamefont
  {Garc\'{i}a-Etxarri}}, \bibinfo {author} {\bibfnamefont {R.}~\bibnamefont
  {G\'{o}mez-Medina}}, \bibinfo {author} {\bibfnamefont {L.~S.}\ \bibnamefont
  {Froufe-P\'{e}rez}}, \bibinfo {author} {\bibfnamefont {C.}~\bibnamefont
  {L\'{o}pez}}, \bibinfo {author} {\bibfnamefont {L.}~\bibnamefont {Chantada}},
  \bibinfo {author} {\bibfnamefont {F.}~\bibnamefont {Scheffold}}, \bibinfo
  {author} {\bibfnamefont {J.}~\bibnamefont {Aizpurua}}, \bibinfo {author}
  {\bibfnamefont {M.}~\bibnamefont {Nieto-Vesperinas}},\ and\ \bibinfo {author}
  {\bibfnamefont {J.~J.}\ \bibnamefont {S\'{a}enz}},\ }\bibfield  {title}
  {\bibinfo {title} {{Strong magnetic response of submicron silicon particles
  in the infrared}},\ }\href@noop {} {\bibfield  {journal} {\bibinfo  {journal}
  {Optics Express}\ }\textbf {\bibinfo {volume} {19}},\ \bibinfo {pages} {4815}
  (\bibinfo {year} {2011})}\BibitemShut {NoStop}%
\bibitem [{\citenamefont {Kuznetsov}\ \emph {et~al.}(2012)\citenamefont
  {Kuznetsov}, \citenamefont {Miroshnichenko}, \citenamefont {Fu},
  \citenamefont {Zhang},\ and\ \citenamefont
  {Luk’yanchuk}}]{AIKuznetsov2012}%
  \BibitemOpen
  \bibfield  {author} {\bibinfo {author} {\bibfnamefont {A.~I.}\ \bibnamefont
  {Kuznetsov}}, \bibinfo {author} {\bibfnamefont {A.~E.}\ \bibnamefont
  {Miroshnichenko}}, \bibinfo {author} {\bibfnamefont {Y.~H.}\ \bibnamefont
  {Fu}}, \bibinfo {author} {\bibfnamefont {J.~B.}\ \bibnamefont {Zhang}},\ and\
  \bibinfo {author} {\bibfnamefont {B.}~\bibnamefont {Luk’yanchuk}},\
  }\bibfield  {title} {\bibinfo {title} {{Magnetic light}},\ }\href@noop {}
  {\bibfield  {journal} {\bibinfo  {journal} {Sci. Rep.}\ }\textbf {\bibinfo
  {volume} {2}},\ \bibinfo {pages} {492} (\bibinfo {year} {2012})}\BibitemShut
  {NoStop}%
\bibitem [{\citenamefont {Picardi}\ \emph {et~al.}(2018)\citenamefont
  {Picardi}, \citenamefont {Zayats},\ and\ \citenamefont
  {Rodr\'{i}guez-Fortu\~{n}o}}]{MFPicardiPRL2018}%
  \BibitemOpen
  \bibfield  {author} {\bibinfo {author} {\bibfnamefont {M.~F.}\ \bibnamefont
  {Picardi}}, \bibinfo {author} {\bibfnamefont {A.~V.}\ \bibnamefont
  {Zayats}},\ and\ \bibinfo {author} {\bibfnamefont {F.~J.}\ \bibnamefont
  {Rodr\'{i}guez-Fortu\~{n}o}},\ }\bibfield  {title} {\bibinfo {title} {{Janus
  and Huygens Dipoles: Near-Field Directionality Beyond Spin-Momentum
  Locking}},\ }\href@noop {} {\bibfield  {journal} {\bibinfo  {journal}
  {Physical Review Letters}\ }\textbf {\bibinfo {volume} {120}},\ \bibinfo
  {pages} {117402} (\bibinfo {year} {2018})}\BibitemShut {NoStop}%
\bibitem [{\citenamefont {Wei}\ and\ \citenamefont
  {Rodr\'{i}guez-Fortu\~{n}o}(2020)}]{LWeiPRAppl2020}%
  \BibitemOpen
  \bibfield  {author} {\bibinfo {author} {\bibfnamefont {L.}~\bibnamefont
  {Wei}}\ and\ \bibinfo {author} {\bibfnamefont {F.~J.}\ \bibnamefont
  {Rodr\'{i}guez-Fortu\~{n}o}},\ }\bibfield  {title} {\bibinfo {title}
  {{Momentum-Space Geometric Structure of Helical Evanescent Waves and Its
  Implications on Near-Field Directionality}},\ }\href@noop {} {\bibfield
  {journal} {\bibinfo  {journal} {Physical Review Applied}\ }\textbf {\bibinfo
  {volume} {13}},\ \bibinfo {pages} {014008} (\bibinfo {year}
  {2020})}\BibitemShut {NoStop}%
\bibitem [{\citenamefont {Geffrin}\ \emph {et~al.}(2012)\citenamefont
  {Geffrin}, \citenamefont {Garc\'{i}a-C\'{a}mara}, \citenamefont
  {G\'{o}mez-Medina}, \citenamefont {Albella}, \citenamefont
  {Froufe-P\'{e}rez}, \citenamefont {Eyraud}, \citenamefont {Litman},
  \citenamefont {Vaillon}, \citenamefont {Gonz\'{a}lez}, \citenamefont
  {Nieto-Vesperinas}, \citenamefont {S\'{a}enz},\ and\ \citenamefont
  {Moreno}}]{JMGeffrinNC2012}%
  \BibitemOpen
  \bibfield  {author} {\bibinfo {author} {\bibfnamefont {J.~M.}\ \bibnamefont
  {Geffrin}}, \bibinfo {author} {\bibfnamefont {B.}~\bibnamefont
  {Garc\'{i}a-C\'{a}mara}}, \bibinfo {author} {\bibfnamefont {R.}~\bibnamefont
  {G\'{o}mez-Medina}}, \bibinfo {author} {\bibfnamefont {P.}~\bibnamefont
  {Albella}}, \bibinfo {author} {\bibfnamefont {L.~S.}\ \bibnamefont
  {Froufe-P\'{e}rez}}, \bibinfo {author} {\bibfnamefont {C.}~\bibnamefont
  {Eyraud}}, \bibinfo {author} {\bibfnamefont {A.}~\bibnamefont {Litman}},
  \bibinfo {author} {\bibfnamefont {R.}~\bibnamefont {Vaillon}}, \bibinfo
  {author} {\bibfnamefont {F.}~\bibnamefont {Gonz\'{a}lez}}, \bibinfo {author}
  {\bibfnamefont {M.}~\bibnamefont {Nieto-Vesperinas}}, \bibinfo {author}
  {\bibfnamefont {J.~J.}\ \bibnamefont {S\'{a}enz}},\ and\ \bibinfo {author}
  {\bibfnamefont {F.}~\bibnamefont {Moreno}},\ }\bibfield  {title} {\bibinfo
  {title} {{Magnetic and electric coherence in forward- and back-scattered
  electromagnetic waves by a single dielectric subwavelength sphere}},\
  }\href@noop {} {\bibfield  {journal} {\bibinfo  {journal} {Nature
  Communications}\ }\textbf {\bibinfo {volume} {3}},\ \bibinfo {pages} {1171}
  (\bibinfo {year} {2012})}\BibitemShut {NoStop}%
\bibitem [{\citenamefont {Fu}\ \emph {et~al.}(2013)\citenamefont {Fu},
  \citenamefont {Kuznetsov}, \citenamefont {Miroshnichenko}, \citenamefont
  {Yu},\ and\ \citenamefont {Luk'yanchuk}}]{YHFuNC2013}%
  \BibitemOpen
  \bibfield  {author} {\bibinfo {author} {\bibfnamefont {Y.~H.}\ \bibnamefont
  {Fu}}, \bibinfo {author} {\bibfnamefont {A.~I.}\ \bibnamefont {Kuznetsov}},
  \bibinfo {author} {\bibfnamefont {A.~E.}\ \bibnamefont {Miroshnichenko}},
  \bibinfo {author} {\bibfnamefont {Y.~F.}\ \bibnamefont {Yu}},\ and\ \bibinfo
  {author} {\bibfnamefont {B.}~\bibnamefont {Luk'yanchuk}},\ }\bibfield
  {title} {\bibinfo {title} {{Directional visible light scattering by silicon
  nanoparticles}},\ }\href@noop {} {\bibfield  {journal} {\bibinfo  {journal}
  {Nature Communications}\ }\textbf {\bibinfo {volume} {4}},\ \bibinfo {pages}
  {1527} (\bibinfo {year} {2013})}\BibitemShut {NoStop}%
\bibitem [{\citenamefont {Person}\ \emph {et~al.}(2013)\citenamefont {Person},
  \citenamefont {Jain}, \citenamefont {Lapin}, \citenamefont {S\'{a}enz},
  \citenamefont {Wicks},\ and\ \citenamefont {Novotny}}]{SPersonNL2013}%
  \BibitemOpen
  \bibfield  {author} {\bibinfo {author} {\bibfnamefont {S.}~\bibnamefont
  {Person}}, \bibinfo {author} {\bibfnamefont {M.}~\bibnamefont {Jain}},
  \bibinfo {author} {\bibfnamefont {Z.}~\bibnamefont {Lapin}}, \bibinfo
  {author} {\bibfnamefont {J.~J.}\ \bibnamefont {S\'{a}enz}}, \bibinfo {author}
  {\bibfnamefont {G.}~\bibnamefont {Wicks}},\ and\ \bibinfo {author}
  {\bibfnamefont {L.}~\bibnamefont {Novotny}},\ }\bibfield  {title} {\bibinfo
  {title} {{Demonstration of Zero Optical Backscattering from Single
  Nanoparticles}},\ }\href@noop {} {\bibfield  {journal} {\bibinfo  {journal}
  {Nano Letters}\ }\textbf {\bibinfo {volume} {13}},\ \bibinfo {pages} {1806}
  (\bibinfo {year} {2013})}\BibitemShut {NoStop}%
\bibitem [{\citenamefont {Picardi}\ \emph
  {et~al.}(2019{\natexlab{a}})\citenamefont {Picardi}, \citenamefont
  {Neugebauer}, \citenamefont {Eismann}, \citenamefont {Leuchs}, \citenamefont
  {Banzer}, \citenamefont {Rodr\'{i}guez-Fortu\~{n}o},\ and\ \citenamefont
  {Zayats}}]{MFPicardiLSA2019}%
  \BibitemOpen
  \bibfield  {author} {\bibinfo {author} {\bibfnamefont {M.~F.}\ \bibnamefont
  {Picardi}}, \bibinfo {author} {\bibfnamefont {M.}~\bibnamefont {Neugebauer}},
  \bibinfo {author} {\bibfnamefont {J.~S.}\ \bibnamefont {Eismann}}, \bibinfo
  {author} {\bibfnamefont {G.}~\bibnamefont {Leuchs}}, \bibinfo {author}
  {\bibfnamefont {P.}~\bibnamefont {Banzer}}, \bibinfo {author} {\bibfnamefont
  {F.~J.}\ \bibnamefont {Rodr\'{i}guez-Fortu\~{n}o}},\ and\ \bibinfo {author}
  {\bibfnamefont {A.~V.}\ \bibnamefont {Zayats}},\ }\bibfield  {title}
  {\bibinfo {title} {{Experimental demonstration of linear and spinning Janus
  dipoles for polarisation- and wavelength-selective near-field coupling}},\
  }\href@noop {} {\bibfield  {journal} {\bibinfo  {journal} {Light:
  Science\&Applications}\ }\textbf {\bibinfo {volume} {8}},\ \bibinfo {pages}
  {52} (\bibinfo {year} {2019}{\natexlab{a}})}\BibitemShut {NoStop}%
\bibitem [{\citenamefont {Staude}\ \emph {et~al.}(2013)\citenamefont {Staude},
  \citenamefont {Miroshnichenko}, \citenamefont {Decker} \emph
  {et~al.}}]{IStaudeACSNano2013}%
  \BibitemOpen
  \bibfield  {author} {\bibinfo {author} {\bibfnamefont {I.}~\bibnamefont
  {Staude}}, \bibinfo {author} {\bibfnamefont {A.~E.}\ \bibnamefont
  {Miroshnichenko}}, \bibinfo {author} {\bibfnamefont {M.}~\bibnamefont
  {Decker}}, \emph {et~al.},\ }\bibfield  {title} {\bibinfo {title} {{Tailoring
  Directional Scattering through Magnetic and Electric Resonances in
  Subwavelength Silicon Nanodisks}},\ }\href@noop {} {\bibfield  {journal}
  {\bibinfo  {journal} {ACS Nano}\ }\textbf {\bibinfo {volume} {7}},\ \bibinfo
  {pages} {7824} (\bibinfo {year} {2013})}\BibitemShut {NoStop}%
\bibitem [{\citenamefont {Pfeiffer}\ and\ \citenamefont
  {Grbic}(2013)}]{CPfeifferPRL2013}%
  \BibitemOpen
  \bibfield  {author} {\bibinfo {author} {\bibfnamefont {C.}~\bibnamefont
  {Pfeiffer}}\ and\ \bibinfo {author} {\bibfnamefont {A.}~\bibnamefont
  {Grbic}},\ }\bibfield  {title} {\bibinfo {title} {{Metamaterial Huygens’
  Surfaces: Tailoring Wave Fronts with Reflectionless Sheets}},\ }\href@noop {}
  {\bibfield  {journal} {\bibinfo  {journal} {Physical Review Letters}\
  }\textbf {\bibinfo {volume} {110}},\ \bibinfo {pages} {197401} (\bibinfo
  {year} {2013})}\BibitemShut {NoStop}%
\bibitem [{\citenamefont {Wei}\ \emph {et~al.}(2018)\citenamefont {Wei},
  \citenamefont {Zayats},\ and\ \citenamefont
  {Rodr\'{i}guez-Fortu\~{n}o}}]{LWeiPRL2018}%
  \BibitemOpen
  \bibfield  {author} {\bibinfo {author} {\bibfnamefont {L.}~\bibnamefont
  {Wei}}, \bibinfo {author} {\bibfnamefont {A.~V.}\ \bibnamefont {Zayats}},\
  and\ \bibinfo {author} {\bibfnamefont {F.~J.}\ \bibnamefont
  {Rodr\'{i}guez-Fortu\~{n}o}},\ }\bibfield  {title} {\bibinfo {title}
  {{Interferometric Evanescent Wave Excitation of a Nanoantenna for
  Ultrasensitive Displacement and Phase Metrology}},\ }\href@noop {} {\bibfield
   {journal} {\bibinfo  {journal} {Physical Review Letters}\ }\textbf {\bibinfo
  {volume} {121}},\ \bibinfo {pages} {193901} (\bibinfo {year}
  {2018})}\BibitemShut {NoStop}%
\bibitem [{\citenamefont {Rodr\'{i}guez-Fortu\~{n}o}\ \emph
  {et~al.}(2013)\citenamefont {Rodr\'{i}guez-Fortu\~{n}o}, \citenamefont
  {Giuseppe}, \citenamefont {Ginzburg}, \citenamefont {O'Connor}, \citenamefont
  {Mart\'{i}nez}, \citenamefont {Wurtz},\ and\ \citenamefont
  {Zayats}}]{FJRFortunoSci2013}%
  \BibitemOpen
  \bibfield  {author} {\bibinfo {author} {\bibfnamefont {F.~J.}\ \bibnamefont
  {Rodr\'{i}guez-Fortu\~{n}o}}, \bibinfo {author} {\bibfnamefont
  {M.}~\bibnamefont {Giuseppe}}, \bibinfo {author} {\bibfnamefont
  {P.}~\bibnamefont {Ginzburg}}, \bibinfo {author} {\bibfnamefont
  {D.}~\bibnamefont {O'Connor}}, \bibinfo {author} {\bibfnamefont
  {A.}~\bibnamefont {Mart\'{i}nez}}, \bibinfo {author} {\bibfnamefont {G.~A.}\
  \bibnamefont {Wurtz}},\ and\ \bibinfo {author} {\bibfnamefont {A.~V.}\
  \bibnamefont {Zayats}},\ }\bibfield  {title} {\bibinfo {title} {{Near-Field
  Interference for the Unidirectional Excitation of Electromagnetic Guided
  Modes}},\ }\href@noop {} {\bibfield  {journal} {\bibinfo  {journal}
  {Science}\ }\textbf {\bibinfo {volume} {340}},\ \bibinfo {pages} {328}
  (\bibinfo {year} {2013})}\BibitemShut {NoStop}%
\bibitem [{\citenamefont {Bliokh}\ \emph
  {et~al.}(2015{\natexlab{a}})\citenamefont {Bliokh}, \citenamefont
  {Rodr\'{i}guez-Fortu\~{n}o}, \citenamefont {Nori},\ and\ \citenamefont
  {Zayats}}]{FJRFortunoNPhoton2015}%
  \BibitemOpen
  \bibfield  {author} {\bibinfo {author} {\bibfnamefont {K.~Y.}\ \bibnamefont
  {Bliokh}}, \bibinfo {author} {\bibfnamefont {F.~J.}\ \bibnamefont
  {Rodr\'{i}guez-Fortu\~{n}o}}, \bibinfo {author} {\bibfnamefont
  {F.}~\bibnamefont {Nori}},\ and\ \bibinfo {author} {\bibfnamefont {A.~V.}\
  \bibnamefont {Zayats}},\ }\bibfield  {title} {\bibinfo {title} {{Spin-orbit
  interactions of light}},\ }\href@noop {} {\bibfield  {journal} {\bibinfo
  {journal} {Nature Photonics}\ }\textbf {\bibinfo {volume} {9}},\ \bibinfo
  {pages} {796} (\bibinfo {year} {2015}{\natexlab{a}})}\BibitemShut {NoStop}%
\bibitem [{\citenamefont {van Mechelen}\ and\ \citenamefont
  {Jacob}(2016)}]{ZJacobOptica2016}%
  \BibitemOpen
  \bibfield  {author} {\bibinfo {author} {\bibfnamefont {T.}~\bibnamefont {van
  Mechelen}}\ and\ \bibinfo {author} {\bibfnamefont {Z.}~\bibnamefont
  {Jacob}},\ }\bibfield  {title} {\bibinfo {title} {{Universal spin-momentum
  locking of evanescent waves}},\ }\href@noop {} {\bibfield  {journal}
  {\bibinfo  {journal} {Optica}\ }\textbf {\bibinfo {volume} {3}},\ \bibinfo
  {pages} {118} (\bibinfo {year} {2016})}\BibitemShut {NoStop}%
\bibitem [{\citenamefont {Bliokh}\ \emph
  {et~al.}(2015{\natexlab{b}})\citenamefont {Bliokh}, \citenamefont
  {Smirnova},\ and\ \citenamefont {Nori}}]{KYBliokhSci2015}%
  \BibitemOpen
  \bibfield  {author} {\bibinfo {author} {\bibfnamefont {K.~Y.}\ \bibnamefont
  {Bliokh}}, \bibinfo {author} {\bibfnamefont {D.}~\bibnamefont {Smirnova}},\
  and\ \bibinfo {author} {\bibfnamefont {F.}~\bibnamefont {Nori}},\ }\bibfield
  {title} {\bibinfo {title} {{Quantum spin Hall effect of light}},\ }\href@noop
  {} {\bibfield  {journal} {\bibinfo  {journal} {Science}\ }\textbf {\bibinfo
  {volume} {348}},\ \bibinfo {pages} {1448} (\bibinfo {year}
  {2015}{\natexlab{b}})}\BibitemShut {NoStop}%
\bibitem [{\citenamefont {Long}\ \emph {et~al.}(2018)\citenamefont {Long},
  \citenamefont {Ren},\ and\ \citenamefont {Chen}}]{YLongPNAS2018}%
  \BibitemOpen
  \bibfield  {author} {\bibinfo {author} {\bibfnamefont {Y.}~\bibnamefont
  {Long}}, \bibinfo {author} {\bibfnamefont {J.}~\bibnamefont {Ren}},\ and\
  \bibinfo {author} {\bibfnamefont {H.}~\bibnamefont {Chen}},\ }\bibfield
  {title} {\bibinfo {title} {{Intrinsic spin of elastic waves}},\ }\href@noop
  {} {\bibfield  {journal} {\bibinfo  {journal} {PNAS}\ }\textbf {\bibinfo
  {volume} {115}},\ \bibinfo {pages} {9951} (\bibinfo {year}
  {2018})}\BibitemShut {NoStop}%
\bibitem [{\citenamefont {Shi}\ \emph {et~al.}(2019)\citenamefont {Shi},
  \citenamefont {Zhao}, \citenamefont {Long}, \citenamefont {Yang},
  \citenamefont {Wang}, \citenamefont {Chen}, \citenamefont {Ren},\ and\
  \citenamefont {Zhang}}]{CShiNSR2019}%
  \BibitemOpen
  \bibfield  {author} {\bibinfo {author} {\bibfnamefont {C.}~\bibnamefont
  {Shi}}, \bibinfo {author} {\bibfnamefont {R.}~\bibnamefont {Zhao}}, \bibinfo
  {author} {\bibfnamefont {Y.}~\bibnamefont {Long}}, \bibinfo {author}
  {\bibfnamefont {S.}~\bibnamefont {Yang}}, \bibinfo {author} {\bibfnamefont
  {Y.}~\bibnamefont {Wang}}, \bibinfo {author} {\bibfnamefont {H.}~\bibnamefont
  {Chen}}, \bibinfo {author} {\bibfnamefont {J.}~\bibnamefont {Ren}},\ and\
  \bibinfo {author} {\bibfnamefont {X.}~\bibnamefont {Zhang}},\ }\bibfield
  {title} {\bibinfo {title} {{Observation of acoustic spin}},\ }\href@noop {}
  {\bibfield  {journal} {\bibinfo  {journal} {National Science Review}\
  }\textbf {\bibinfo {volume} {6}},\ \bibinfo {pages} {707} (\bibinfo {year}
  {2019})}\BibitemShut {NoStop}%
\bibitem [{\citenamefont {Bliokh}\ and\ \citenamefont
  {Nori}(2019)}]{KYBliokhPRB2018R}%
  \BibitemOpen
  \bibfield  {author} {\bibinfo {author} {\bibfnamefont {K.~Y.}\ \bibnamefont
  {Bliokh}}\ and\ \bibinfo {author} {\bibfnamefont {F.}~\bibnamefont {Nori}},\
  }\bibfield  {title} {\bibinfo {title} {{Transverse spin and surface waves in
  acoustic metamaterials}},\ }\href@noop {} {\bibfield  {journal} {\bibinfo
  {journal} {Physical Review B}\ }\textbf {\bibinfo {volume} {99}},\ \bibinfo
  {pages} {020301(R)} (\bibinfo {year} {2019})}\BibitemShut {NoStop}%
\bibitem [{\citenamefont {Toftul}\ \emph {et~al.}(2019)\citenamefont {Toftul},
  \citenamefont {Bliokh}, \citenamefont {Petrov},\ and\ \citenamefont
  {Nori}}]{IDToftulPRL2019}%
  \BibitemOpen
  \bibfield  {author} {\bibinfo {author} {\bibfnamefont {I.~D.}\ \bibnamefont
  {Toftul}}, \bibinfo {author} {\bibfnamefont {K.~Y.}\ \bibnamefont {Bliokh}},
  \bibinfo {author} {\bibfnamefont {M.~I.}\ \bibnamefont {Petrov}},\ and\
  \bibinfo {author} {\bibfnamefont {F.}~\bibnamefont {Nori}},\ }\bibfield
  {title} {\bibinfo {title} {{Acoustic Radiation Force and Torque on Small
  Particles as Measures of the Canonical Momentum and Spin Densities}},\
  }\href@noop {} {\bibfield  {journal} {\bibinfo  {journal} {Physical Review
  Letters}\ }\textbf {\bibinfo {volume} {123}},\ \bibinfo {pages} {183901}
  (\bibinfo {year} {2019})}\BibitemShut {NoStop}%
\bibitem [{\citenamefont {Burns}\ \emph {et~al.}()\citenamefont {Burns},
  \citenamefont {Bliokh}, \citenamefont {Nori},\ and\ \citenamefont
  {Dressel}}]{LBurnsArXiv2019}%
  \BibitemOpen
  \bibfield  {author} {\bibinfo {author} {\bibfnamefont {L.}~\bibnamefont
  {Burns}}, \bibinfo {author} {\bibfnamefont {K.~Y.}\ \bibnamefont {Bliokh}},
  \bibinfo {author} {\bibfnamefont {F.}~\bibnamefont {Nori}},\ and\ \bibinfo
  {author} {\bibfnamefont {J.}~\bibnamefont {Dressel}},\ }\bibfield  {title}
  {\bibinfo {title} {{Acoustic field theory: scalar, vector, spinor
  representations and the emergence of acoustic spin}},\ }\href@noop {}
  {\bibinfo  {journal} {arXiv:1912.10522}\ }\BibitemShut {NoStop}%
\bibitem [{\citenamefont {Golat}\ \emph {et~al.}()\citenamefont {Golat},
  \citenamefont {Lim},\ and\ \citenamefont
  {Rodr\'{i}guez-Fortu\~{n}o}}]{SGolatArxiv2019}%
  \BibitemOpen
\bibfield  {journal} {  }\bibfield  {author} {\bibinfo {author} {\bibfnamefont
  {S.}~\bibnamefont {Golat}}, \bibinfo {author} {\bibfnamefont {E.~A.}\
  \bibnamefont {Lim}},\ and\ \bibinfo {author} {\bibfnamefont {F.~J.}\
  \bibnamefont {Rodr\'{i}guez-Fortu\~{n}o}},\ }\bibfield  {title} {\bibinfo
  {title} {{Evanescent Gravitational Waves}},\ }\href@noop {} {\bibinfo
  {journal} {arXiv:1903.09690}\ }\BibitemShut {NoStop}%
\bibitem [{\citenamefont {Kafesaki}\ \emph {et~al.}(2000)\citenamefont
  {Kafesaki}, \citenamefont {Penciu},\ and\ \citenamefont
  {Economou}}]{MKafesakiPRL2000}%
  \BibitemOpen
\bibfield  {journal} {  }\bibfield  {author} {\bibinfo {author} {\bibfnamefont
  {M.}~\bibnamefont {Kafesaki}}, \bibinfo {author} {\bibfnamefont {R.~S.}\
  \bibnamefont {Penciu}},\ and\ \bibinfo {author} {\bibfnamefont {E.~N.}\
  \bibnamefont {Economou}},\ }\bibfield  {title} {\bibinfo {title} {{Air
  Bubbles in Water: A Strongly Multiple Scattering Medium for Acoustic
  Waves}},\ }\href@noop {} {\bibfield  {journal} {\bibinfo  {journal} {Physical
  Review Letters}\ }\textbf {\bibinfo {volume} {84}},\ \bibinfo {pages} {6050}
  (\bibinfo {year} {2000})}\BibitemShut {NoStop}%
\bibitem [{\citenamefont {Errico}\ \emph {et~al.}(2005)\citenamefont {Errico},
  \citenamefont {Pierre}, \citenamefont {Pezet}, \citenamefont {Dessaily},
  \citenamefont {Lenkei}, \citenamefont {Couture},\ and\ \citenamefont
  {Tanter}}]{CErricoNature2005}%
  \BibitemOpen
  \bibfield  {author} {\bibinfo {author} {\bibfnamefont {C.}~\bibnamefont
  {Errico}}, \bibinfo {author} {\bibfnamefont {J.}~\bibnamefont {Pierre}},
  \bibinfo {author} {\bibfnamefont {S.}~\bibnamefont {Pezet}}, \bibinfo
  {author} {\bibfnamefont {Y.}~\bibnamefont {Dessaily}}, \bibinfo {author}
  {\bibfnamefont {Z.}~\bibnamefont {Lenkei}}, \bibinfo {author} {\bibfnamefont
  {O.}~\bibnamefont {Couture}},\ and\ \bibinfo {author} {\bibfnamefont
  {M.}~\bibnamefont {Tanter}},\ }\bibfield  {title} {\bibinfo {title}
  {{Ultrafast ultrasound localization microscopy for deep super-resolution
  vascular imaging}},\ }\href@noop {} {\bibfield  {journal} {\bibinfo
  {journal} {Nature}\ }\textbf {\bibinfo {volume} {527}},\ \bibinfo {pages}
  {499} (\bibinfo {year} {2005})}\BibitemShut {NoStop}%
\bibitem [{\citenamefont {Cummer}\ \emph {et~al.}(2016)\citenamefont {Cummer},
  \citenamefont {Christensen},\ and\ \citenamefont
  {Alu}}]{SACummerNatureRM2016}%
  \BibitemOpen
  \bibfield  {author} {\bibinfo {author} {\bibfnamefont {S.~A.}\ \bibnamefont
  {Cummer}}, \bibinfo {author} {\bibfnamefont {J.}~\bibnamefont
  {Christensen}},\ and\ \bibinfo {author} {\bibfnamefont {A.}~\bibnamefont
  {Alu}},\ }\bibfield  {title} {\bibinfo {title} {{Controlling sound with
  acoustic metamaterials}},\ }\href@noop {} {\bibfield  {journal} {\bibinfo
  {journal} {Nature Reviews Materials}\ }\textbf {\bibinfo {volume} {1}},\
  \bibinfo {pages} {16001} (\bibinfo {year} {2016})}\BibitemShut {NoStop}%
\bibitem [{\citenamefont {Zangeneh-Nejad}\ and\ \citenamefont
  {Fleury}(2019)}]{FZangenehNejadRevPhys2019}%
  \BibitemOpen
  \bibfield  {author} {\bibinfo {author} {\bibfnamefont {F.}~\bibnamefont
  {Zangeneh-Nejad}}\ and\ \bibinfo {author} {\bibfnamefont {R.}~\bibnamefont
  {Fleury}},\ }\bibfield  {title} {\bibinfo {title} {{Active times for acoustic
  metamaterials}},\ }\href@noop {} {\bibfield  {journal} {\bibinfo  {journal}
  {Reviews in Physics}\ }\textbf {\bibinfo {volume} {4}},\ \bibinfo {pages}
  {100031} (\bibinfo {year} {2019})}\BibitemShut {NoStop}%
\bibitem [{\citenamefont {Ba}\ \emph {et~al.}(2017)\citenamefont {Ba},
  \citenamefont {Kovalenko}, \citenamefont {Arist\'{e}gui}, \citenamefont
  {Mondain-Monval},\ and\ \citenamefont {Brunet}}]{ABaSR2017}%
  \BibitemOpen
  \bibfield  {author} {\bibinfo {author} {\bibfnamefont {A.}~\bibnamefont
  {Ba}}, \bibinfo {author} {\bibfnamefont {A.}~\bibnamefont {Kovalenko}},
  \bibinfo {author} {\bibfnamefont {C.}~\bibnamefont {Arist\'{e}gui}}, \bibinfo
  {author} {\bibfnamefont {O.}~\bibnamefont {Mondain-Monval}},\ and\ \bibinfo
  {author} {\bibfnamefont {T.}~\bibnamefont {Brunet}},\ }\bibfield  {title}
  {\bibinfo {title} {{Soft porous silicone rubbers with ultra-low sound speeds
  in acoustic metamaterials}},\ }\href@noop {} {\bibfield  {journal} {\bibinfo
  {journal} {Scientific Reports}\ }\textbf {\bibinfo {volume} {7}},\ \bibinfo
  {pages} {40106} (\bibinfo {year} {2017})}\BibitemShut {NoStop}%
\bibitem [{\citenamefont {Li}\ and\ \citenamefont
  {Chan}(2004)}]{JensenLiPRE2004}%
  \BibitemOpen
  \bibfield  {author} {\bibinfo {author} {\bibfnamefont {J.}~\bibnamefont
  {Li}}\ and\ \bibinfo {author} {\bibfnamefont {C.~T.}\ \bibnamefont {Chan}},\
  }\bibfield  {title} {\bibinfo {title} {{Double-negative acoustic
  metamaterial}},\ }\href@noop {} {\bibfield  {journal} {\bibinfo  {journal}
  {Physical Review E}\ }\textbf {\bibinfo {volume} {70}},\ \bibinfo {pages}
  {055602(R)} (\bibinfo {year} {2004})}\BibitemShut {NoStop}%
\bibitem [{\citenamefont {Ding}\ \emph {et~al.}(2007)\citenamefont {Ding},
  \citenamefont {Liu}, \citenamefont {Qiu},\ and\ \citenamefont
  {Shi}}]{YDingPRL2007}%
  \BibitemOpen
  \bibfield  {author} {\bibinfo {author} {\bibfnamefont {Y.}~\bibnamefont
  {Ding}}, \bibinfo {author} {\bibfnamefont {Z.}~\bibnamefont {Liu}}, \bibinfo
  {author} {\bibfnamefont {C.}~\bibnamefont {Qiu}},\ and\ \bibinfo {author}
  {\bibfnamefont {J.}~\bibnamefont {Shi}},\ }\bibfield  {title} {\bibinfo
  {title} {{Metamaterial with Simultaneously Negative Bulk Modulus and Mass
  Density}},\ }\href@noop {} {\bibfield  {journal} {\bibinfo  {journal}
  {Physical Review Letters}\ }\textbf {\bibinfo {volume} {99}},\ \bibinfo
  {pages} {093904} (\bibinfo {year} {2007})}\BibitemShut {NoStop}%
\bibitem [{\citenamefont {Jordaan}\ \emph {et~al.}(2018)\citenamefont
  {Jordaan}, \citenamefont {Punzet}, \citenamefont {Melnikov}, \citenamefont
  {Sanches}, \citenamefont {Oberst}, \citenamefont {Marburg},\ and\
  \citenamefont {Powell}}]{JJordaanAPL2018}%
  \BibitemOpen
  \bibfield  {author} {\bibinfo {author} {\bibfnamefont {J.}~\bibnamefont
  {Jordaan}}, \bibinfo {author} {\bibfnamefont {S.}~\bibnamefont {Punzet}},
  \bibinfo {author} {\bibfnamefont {A.}~\bibnamefont {Melnikov}}, \bibinfo
  {author} {\bibfnamefont {A.}~\bibnamefont {Sanches}}, \bibinfo {author}
  {\bibfnamefont {S.}~\bibnamefont {Oberst}}, \bibinfo {author} {\bibfnamefont
  {S.}~\bibnamefont {Marburg}},\ and\ \bibinfo {author} {\bibfnamefont {D.~A.}\
  \bibnamefont {Powell}},\ }\bibfield  {title} {\bibinfo {title} {{Measuring
  monopole and dipole polarizability of acoustic meta-atoms}},\ }\href@noop {}
  {\bibfield  {journal} {\bibinfo  {journal} {Applied Physics Letters}\
  }\textbf {\bibinfo {volume} {113}},\ \bibinfo {pages} {224102} (\bibinfo
  {year} {2018})}\BibitemShut {NoStop}%
\bibitem [{\citenamefont {Lu}\ \emph {et~al.}(2017)\citenamefont {Lu},
  \citenamefont {Ding}, \citenamefont {Wang}, \citenamefont {Peng},
  \citenamefont {Cui}, \citenamefont {Liu},\ and\ \citenamefont
  {Liu}}]{GLuAPL2017}%
  \BibitemOpen
  \bibfield  {author} {\bibinfo {author} {\bibfnamefont {G.}~\bibnamefont
  {Lu}}, \bibinfo {author} {\bibfnamefont {E.}~\bibnamefont {Ding}}, \bibinfo
  {author} {\bibfnamefont {Y.}~\bibnamefont {Wang}}, \bibinfo {author}
  {\bibfnamefont {X.}~\bibnamefont {Peng}}, \bibinfo {author} {\bibfnamefont
  {J.}~\bibnamefont {Cui}}, \bibinfo {author} {\bibfnamefont {X.}~\bibnamefont
  {Liu}},\ and\ \bibinfo {author} {\bibfnamefont {X.}~\bibnamefont {Liu}},\
  }\bibfield  {title} {\bibinfo {title} {{Realization of acoustic wave
  directivity at low frequencies with a subwavelength Mie resonant
  structure}},\ }\href@noop {} {\bibfield  {journal} {\bibinfo  {journal}
  {Applied Physics Letters}\ }\textbf {\bibinfo {volume} {110}},\ \bibinfo
  {pages} {123507} (\bibinfo {year} {2017})}\BibitemShut {NoStop}%
\bibitem [{\citenamefont {Cheng}\ \emph {et~al.}(2015)\citenamefont {Cheng},
  \citenamefont {Zhou}, \citenamefont {Yuan}, \citenamefont {Wu}, \citenamefont
  {Wei},\ and\ \citenamefont {Liu}}]{YChengNatMat2015}%
  \BibitemOpen
  \bibfield  {author} {\bibinfo {author} {\bibfnamefont {Y.}~\bibnamefont
  {Cheng}}, \bibinfo {author} {\bibfnamefont {C.}~\bibnamefont {Zhou}},
  \bibinfo {author} {\bibfnamefont {B.~G.}\ \bibnamefont {Yuan}}, \bibinfo
  {author} {\bibfnamefont {D.~J.}\ \bibnamefont {Wu}}, \bibinfo {author}
  {\bibfnamefont {Q.}~\bibnamefont {Wei}},\ and\ \bibinfo {author}
  {\bibfnamefont {X.~J.}\ \bibnamefont {Liu}},\ }\bibfield  {title} {\bibinfo
  {title} {{Ultra-sparse metasurface for high reflection of low-frequency sound
  based on artificial Mie resonances}},\ }\href@noop {} {\bibfield  {journal}
  {\bibinfo  {journal} {Nature Materials}\ }\textbf {\bibinfo {volume} {14}},\
  \bibinfo {pages} {1013} (\bibinfo {year} {2015})}\BibitemShut {NoStop}%
\bibitem [{\citenamefont {Novotny}(1997)}]{LNovotnyJOSAA1997}%
  \BibitemOpen
  \bibfield  {author} {\bibinfo {author} {\bibfnamefont {L.}~\bibnamefont
  {Novotny}},\ }\bibfield  {title} {\bibinfo {title} {{Allowed and forbidden
  light in near-field optics. I. A single dipolar light source}},\ }\href@noop
  {} {\bibfield  {journal} {\bibinfo  {journal} {J. Opt. Soc. Am. A}\ }\textbf
  {\bibinfo {volume} {14}},\ \bibinfo {pages} {91} (\bibinfo {year}
  {1997})}\BibitemShut {NoStop}%
\bibitem [{\citenamefont {Picardi}\ \emph {et~al.}(2017)\citenamefont
  {Picardi}, \citenamefont {Manjavacas}, \citenamefont {Zayats},\ and\
  \citenamefont {Rodr\'{i}guez-Fortu\~{n}o}}]{MFPicardiPRB2017}%
  \BibitemOpen
  \bibfield  {author} {\bibinfo {author} {\bibfnamefont {M.~F.}\ \bibnamefont
  {Picardi}}, \bibinfo {author} {\bibfnamefont {A.}~\bibnamefont {Manjavacas}},
  \bibinfo {author} {\bibfnamefont {A.~V.}\ \bibnamefont {Zayats}},\ and\
  \bibinfo {author} {\bibfnamefont {F.~J.}\ \bibnamefont
  {Rodr\'{i}guez-Fortu\~{n}o}},\ }\bibfield  {title} {\bibinfo {title}
  {{Unidirectional evanescent-wave coupling from circularly polarized electric
  and magnetic dipoles: An angular spectrum approach}},\ }\href@noop {}
  {\bibfield  {journal} {\bibinfo  {journal} {Physical Review B}\ }\textbf
  {\bibinfo {volume} {95}},\ \bibinfo {pages} {245416} (\bibinfo {year}
  {2017})}\BibitemShut {NoStop}%
\bibitem [{\citenamefont {Mandel}\ and\ \citenamefont
  {Wolf}(1995)}]{LMandelBOOK}%
  \BibitemOpen
  \bibfield  {author} {\bibinfo {author} {\bibfnamefont {L.}~\bibnamefont
  {Mandel}}\ and\ \bibinfo {author} {\bibfnamefont {E.}~\bibnamefont {Wolf}},\
  }\href@noop {} {\emph {\bibinfo {title} {{Optical Coherence and Quantum
  Optics}}}}\ (\bibinfo  {publisher} {Cambridge University Press},\ \bibinfo
  {address} {Cambridge, UK},\ \bibinfo {year} {1995})\BibitemShut {NoStop}%
\bibitem [{\citenamefont {Nieto-Vesperinas}(2006)}]{MNietoVesperinasBOOK}%
  \BibitemOpen
  \bibfield  {author} {\bibinfo {author} {\bibfnamefont {M.}~\bibnamefont
  {Nieto-Vesperinas}},\ }\href@noop {} {\emph {\bibinfo {title} {{Scattering
  and Diffraction in Physical Optics}}}}\ (\bibinfo  {publisher} {World
  Scientific},\ \bibinfo {address} {Singapore; Hackensack, NJ},\ \bibinfo
  {year} {2006})\BibitemShut {NoStop}%
\bibitem [{\citenamefont {Maynard}\ \emph {et~al.}(1985)\citenamefont
  {Maynard}, \citenamefont {Williams},\ and\ \citenamefont
  {Lee}}]{JDMaynardJASA1985}%
  \BibitemOpen
  \bibfield  {author} {\bibinfo {author} {\bibfnamefont {J.~D.}\ \bibnamefont
  {Maynard}}, \bibinfo {author} {\bibfnamefont {E.~G.}\ \bibnamefont
  {Williams}},\ and\ \bibinfo {author} {\bibfnamefont {Y.}~\bibnamefont
  {Lee}},\ }\bibfield  {title} {\bibinfo {title} {{Nearfield acoustic
  holography: I. Theory of generalized holography and the development of
  NAH}},\ }\href@noop {} {\bibfield  {journal} {\bibinfo  {journal} {Journal of
  the Acoustical Society of America}\ }\textbf {\bibinfo {volume} {78}},\
  \bibinfo {pages} {1395} (\bibinfo {year} {1985})}\BibitemShut {NoStop}%
\bibitem [{\citenamefont {Picardi}\ \emph
  {et~al.}(2019{\natexlab{b}})\citenamefont {Picardi}, \citenamefont {Zayats},\
  and\ \citenamefont {Rodr\'{i}guez-Fortu\~{n}o}}]{MFPicardiLPR2019}%
  \BibitemOpen
  \bibfield  {author} {\bibinfo {author} {\bibfnamefont {M.~F.}\ \bibnamefont
  {Picardi}}, \bibinfo {author} {\bibfnamefont {A.~V.}\ \bibnamefont
  {Zayats}},\ and\ \bibinfo {author} {\bibfnamefont {F.~J.}\ \bibnamefont
  {Rodr\'{i}guez-Fortu\~{n}o}},\ }\bibfield  {title} {\bibinfo {title}
  {{Amplitude and Phase Control of Guided Modes Excitation from a Single Dipole
  Source: Engineering Far‐and Near‐Field Directionality}},\ }\href@noop {}
  {\bibfield  {journal} {\bibinfo  {journal} {Laser and Photonics Review}\
  }\textbf {\bibinfo {volume} {13}},\ \bibinfo {pages} {1900250} (\bibinfo
  {year} {2019}{\natexlab{b}})}\BibitemShut {NoStop}%
\bibitem [{\citenamefont {Long}\ \emph {et~al.}(2020)\citenamefont {Long},
  \citenamefont {Ge}, \citenamefont {Zhang}, \citenamefont {Xu}, \citenamefont
  {Ren}, \citenamefont {Lu}, \citenamefont {Bao}, \citenamefont {Chen},\ and\
  \citenamefont {Chen}}]{YLongNSR2020}%
  \BibitemOpen
  \bibfield  {author} {\bibinfo {author} {\bibfnamefont {Y.}~\bibnamefont
  {Long}}, \bibinfo {author} {\bibfnamefont {H.}~\bibnamefont {Ge}}, \bibinfo
  {author} {\bibfnamefont {D.}~\bibnamefont {Zhang}}, \bibinfo {author}
  {\bibfnamefont {X.}~\bibnamefont {Xu}}, \bibinfo {author} {\bibfnamefont
  {J.}~\bibnamefont {Ren}}, \bibinfo {author} {\bibfnamefont {M.~H.}\
  \bibnamefont {Lu}}, \bibinfo {author} {\bibfnamefont {M.}~\bibnamefont
  {Bao}}, \bibinfo {author} {\bibfnamefont {H.}~\bibnamefont {Chen}},\ and\
  \bibinfo {author} {\bibfnamefont {Y.~F.}\ \bibnamefont {Chen}},\ }\bibfield
  {title} {\bibinfo {title} {{Symmetry Selective Directionality in Near-Field
  Acoustics}},\ }\href@noop {} {\bibfield  {journal} {\bibinfo  {journal}
  {National Science Review}\ ,\ \bibinfo {pages} {nwaa040}} (\bibinfo {year}
  {2020})}\BibitemShut {NoStop}%
\end{thebibliography}%

\appendix

\section{Scattering coefficients of spherical particle and angular spectrum of acoustic monopole and dipole}

In this work, we consider an acoustic scatterer with dominant monopolar and dipolar responses, subject to an external time-harmonic
sound wave with a pressure distribution $p_{\mathrm{in}}(\mathbf{r},t)=p_{\mathrm{in}}(\mathbf{r})\mathrm{e}^{-i\omega t}$ and $p_{\mathrm{in}}(\mathbf{r})= p_0 \mathrm{e}^{i\mathbf{k_{0}\mathbf{r}}}$. We assume that the scatterer is located at the origin $\mathbf{r}=0$ in a fluidic background with mass density $\rho_0$ and compressibility $\beta_0$ and only longitudinal sound waves are considered. The longitudinal sound velocity of the background medium is $c_0=1/\sqrt{\rho_0\beta_0}=\omega/k_{0}$ with $k_0=2\pi/\lambda$ being the wave-number of the background medium and $\lambda$ the wavelength of the sound wave.

The contribution of the monopolar and dipolar responses of the acoustic scatterer to the total scattering is purely determined by the monopole moment $M$ and dipole moment $\mathbf{D}$:

\begin{align}
M&=\alpha_{M} p_{\mathrm{in}}(\mathbf{r}=0),\\\nonumber
\mathbf{D}&=\alpha_{D} \sqrt{\frac{\rho_0}{\beta_0}}\mathbf{v}_{\mathrm{in}}(\mathbf{r}=0),
\end{align}\label{eq:MDmoments}

\noindent where $\mathbf{v}_{\mathrm{in}}=\frac{1}{i\omega\rho_0}\nabla p_{\mathrm{in}}$ is the velocity field of the incident sound wave, $\alpha_{M}=a_0/i$ and $\alpha_{D}=3a_1/i$ represent the monopole and dipole strength of the scatterer, and where $a_0$ and $a_1$ are coefficients for the monopole and dipole, solely determined by the scatterer itself. For the special case of a spherical object, its scattering of sound waves can be analytically treated like Mie theory for optical scattering. Consider a spherical scatterer with a radius of $r_0$ and made of materials with mass density $\rho_1$ and compressibility $\beta_1$, supporting longitudinal sound waves with a velocity $c_1=1/\sqrt{\rho_1\beta_1}$. The coefficients $a_0$ and $a_1$ can be determined by the following expression:

\begin{equation}\label{eq:AcousticMie}
a_n=\frac{\sqrt{\bar{\beta}/\bar{\rho}}j^{'}_n(k_1r_0)j_n(k_0r_0)-j_n(k_1r_0)j^{'}_n(k_0r_0)}{j_n(k_1r_0)h^{(1)'}_n(k_0r_0)-\sqrt{\bar{\beta}/\bar{\rho}}j^{'}_n(k_1r_0)h^{(1)}_n(k_0r_0)},
\end{equation}

\noindent where $j_n(kr)$ is the spherical Bessel function of the first kind and $h^{(1)}_n(kr)$ is the spherical Hankel function of the first kind, $j^{'}_n(kr)$ and $h^{(1)'}_n(kr)$ are their first order derivatives with respect to the argument variable $kr$. The relative mass density and compressibility are defined as $\bar{\rho}=\rho_1/\rho_0$ and $\bar{\beta}=\beta_1/\beta_0$, and $k_1=\omega/c_1=k_0 n$ with $n=\sqrt{\bar{\rho}\bar{\beta}}$ being the acoustic refractive index.

The scattering pressure distribution due to the monopole and dipole contribution can be expressed as:

\begin{align}\label{eq:MDfield}
p(\mathbf{r})&=M \frac{\mathrm{e}^{ik_0r}}{k_0r} + \frac{1}{ik_0}\mathbf{D}\cdot\nabla \left(\frac{\mathrm{e}^{ik_0r}}{k_0r}\right).
\end{align}

In order to obtain the angular spectrum of this source, we need to perform a partial Fourier transform in the $xy$ plane. The angular spectrum $p(k_x,k_y)$ is defined such that:

\begin{equation}
\label{eq:partialfourier}
p(\mathbf{r})=\iint_{-\infty}^{+\infty} p(k_x,k_y) \mathrm{e}^{i(k_xx+k_yy+k_z|z|)} \mathrm{d}k_x\mathrm{d}k_y,
\end{equation}

\noindent where the $z$ direction is an arbitrarily defined direction in space, so that $k_z$ is taken to be the dependent variable $k_z=(k_0^2 - k_x^2 - k_y^2)^{1/2}$ while the angular spectrum is defined in the two dimensional domain of transverse wave-vectors $p(\mathbf{k})=p(k_x,k_y)$. As is well-known in electromagnetism, strictly speaking two different spectra $p^+(k_x,k_y)$ and $p^-(k_x,k_y)$ have to be defined, corresponding to the fields in the $z>0$ halfspace and the $z<0$ halfspace, corresponding to the two possible signs of $k_z$, respectively.

In order to write Eq.\ (\ref{eq:MDfield}) in the form of Eq.\ (\ref{eq:partialfourier}), we make use of Weyl's identity \cite{LMandelBOOK}:

\begin{equation}
\label{eq:Weyl}
\frac{\mathrm{e}^{ik_0r}}{r}=\iint_{-\infty}^{+\infty}\frac{i}{2 \pi k_z}\mathrm{e}^{i(k_xx+k_yy+k_z|z|)}\mathrm{d}k_x\mathrm{d}k_y.
\end{equation}

Substituting Weyl's identity into the corresponding terms in Eq.\ (\ref{eq:MDfield}), and applying the linearity of the integration and gradient operations (the gradient operator becomes a multiplication times $i \mathbf{k}$ inside the integral) we arrive at:

\begin{align}
\label{eq:MDtotalfieldsuppl}
p(\mathbf{r})=\iint_{-\infty}^{+\infty} \frac{i}{2 \pi k_0 k_z}& \left[M+\left(\hat{\mathbf{k}}\cdot\mathbf{D}\right)\right] \\\nonumber
& \mathrm{e}^{i(k_xx+k_yy+k_z|z|)}\mathrm{d}k_x\mathrm{d}k_y,
\end{align}

\noindent where $\hat{\mathbf{k}}=\mathbf{k}/k_0$. By comparing Eq.\ (\ref{eq:MDtotalfieldsuppl}) with the definition of the angular spectrum in Eq.\ (\ref{eq:partialfourier}), we finally identify the expression for the angular spectrum: 

\begin{equation}\label{eq:MDangularspectrum}
p(k_x,k_y)=\frac{i}{2 \pi k_0 k_z}\left[M+\left(\hat{\mathbf{k}}\cdot\mathbf{D}\right)\right],
\end{equation}

\noindent as given in the main text. Note that the two angular spectra $p^+(k_x,k_y)$ and $p^-(k_x,k_y)$, corresponding to the two half spaces $z>0$ and $z<0$ will differ in the sign of the $z$ component of the vector $\hat{\mathbf{k}}$. Also note that Eq.\ (\ref{eq:MDangularspectrum}) is a complete and exact analytical form of the angular spectrum, revealing not only the far-field directionality of the source (for $k_x^2+k_y^2 \leq k_0^2$) but also its near-field directionality (corresponding to the evanescent wave spectrum when $k_x^2+k_y^2 > k_0^2$) associated to the coupling behaviour between this source and nearby bound waveguide modes.
\section{Scattering coefficients of cylindrical particles}
The monopole and dipole strength of a cylindrical scatterer can be determined as $\alpha_{M}=a_0/i$ and $\alpha_{D}=2a_1/i$, with the acoustic Mie coefficients $a_0$ and $a_1$ determined by the following expression:
\begin{equation}\label{eq:AcousticMiecyl}
a_n=\frac{\sqrt{\bar{\beta}/\bar{\rho}}J^{'}_n(k_1r_0)J_n(k_0r_0)-J_n(k_1r_0)J^{'}_n(k_0r_0)}{J_n(k_1r_0)H^{(1)'}_n(k_0r_0)-\sqrt{\bar{\beta}/\bar{\rho}}J^{'}_n(k_1r_0)H^{(1)}_n(k_0r_0)},
\end{equation}
\noindent where $J_n(kr)$ is the Bessel function of the first kind and $H^{(1)}_n(kr)$ is the Hankel function of the first kind, $J^{'}_n(kr)$ and $H^{(1)'}_n(kr)$ are their first order derivatives with respect to the argument variable $kr$.
\end{document}